\documentclass[12pt]{article}
\begin{document}

\begin{titlepage}

\begin{center}
{\large\bf An Operational Approach To Black Hole Entropy}
\end{center}

\vfill

\begin{center}
F. Pretorius, D. Vollick and W. Israel
\end{center}
\vfill

\begin{center}{\sl
Canadian Institute for Advanced Research Cosmology Program\\
Department of Physics and Astronomy \\
University of Victoria \\
P.O. Box 3055 STN CSC \\
Victoria, B.C. V8W 3P6 \\
}\end{center}
\vfill

\begin{abstract} 
In this paper we calculate the entropy of a thin spherical shell that 
contracts reversibly from infinity down to its event horizon. We find that, 
for a broad class of equations of state, the entropy of a non-extremal shell 
is one-quarter of its area in the black hole limit. The considerations in this 
paper suggest the following operational definition for the entropy of a 
black hole: $S_{BH}$ is the equilibrium thermodynamic entropy that 
would be stored in the material which gathers to form the black hole, if all 
of this material were compressed into a thin layer near its gravitational 
radius. Since the entropy for a given mass and area is maximized for 
thermal equilibrium we expect that this is the maximum entropy that could 
be stored in the material before it crosses the horizon. In the case of an 
extremal black hole the shell model does not assign an unambiguous value 
to the entropy. 

\end{abstract}

\end{titlepage}

\section{Introduction}
As is well known \cite{hawking} the classical laws of black hole dynamics 
together with the Hawking temperature 

\begin{equation}
T_H=\frac{\kappa}{2\pi} 
\end{equation} 

lead to the Bekenstein-Hawking postulate that the entropy of a black hole 
is given by

\begin{equation}
S_{BH}=\frac{1}{4}A,
\end{equation}

where $\kappa$ is the surface gravity of the hole and $A$ is its area (here 
we work in units with $G=c=\hbar=1$). This expression for the black hole 
entropy has raised some important questions that still remain unanswered. 
At what stage in the black hole's evolution is its entropy created? Is it 
created immediately upon the formation by gravitational collapse or only 
gradually over the (typically) long course of evaporation? What is the 
dynamical mechanism that makes $S_{BH}$ a universal function, 
independent of the hole's past history and detailed internal condition? 
There are a variety of possible answers. First the subtlest possibility: It is 
conceivable that no quantum entropy is irreversibly created by the hole. 
No information is lost and $S_{BH}$ is merely a measure of our own 
temporary loss of access (during the lifetime of the black hole) to 
correlations beneath the horizon. When the black hole finally evaporates 
these correlations will be fully visible to us. A black hole formed by the 
collapse of matter in a pure state will evaporate into a radiation field 
whose distribution appears thermal on a coarse-grained level, but will be 
recognized to be in a pure state once the last particles have left the hole. 
\par
Alternatively, it is possible that black hole formation and evaporation is 
accompanied by an irreversible increase of entropy, and we come back to 
the questions how, where, and when? The original (pre-1974) motivation 
for assigning an entropy to a black hole was to keep account of the thermal 
entropy of objects thrown into a hole (Wheeler's teacup experiment). The 
generalized second law of black hole thermodynamics, which states that 
$S_{BH}$ plus the external entropy is non-decreasing, lends support to 
this view of black hole entropy. Nevertheless, it is not possible (as 
emphasized by Kundt \cite{kundt} more than 20 years ago) simply to 
identify $S_{BH}$ with the thermal entropy of all the matter which 
collapsed to form the black hole. Since this is an issue of principle, we 
could, for instance, consider an idealized Oppenheimer-Snyder collapse of 
cold, pressureless, viscous-free dust: here, no material entropy ever 
develops. \par
The view that entropy is somehow created in the process of evaporation 
also meets with difficulties. Black hole evaporation is very nearly, and can 
be made exactly, reversible. We simply enclose the hole in a container, so 
that it comes into equilibrium with its own radiation. We then poke a small 
hole in the container and let the radiation leak out arbitrarily slowly. Since 
this process is reversible, no entropy is generated.\par
Another possible explanation of the entropy of a black hole stems from 
Frolov and Novikov \cite{frolov}. This links $S_{BH}$ with modes 
(produced by vacuum fluctuations) propagating ``outwards" just inside and 
alongside the horizon. These modes have positive frequency but their 
energy is negative as calibrated for an observer at infinity (i.e. including 
the contribution of gravitational potential energy). Their spectrum is 
thermal with temperature $T_H$. Detailed implementation of this picture 
is so far still plagued with divergences and ambiguities. \par
Finally we mention that Zurek and Thorne \cite{zurek} have suggested 
that $S_{BH}$ should be interpreted as the logarithm of ``the number of 
quantum-mechanically distinct ways that the black hole could have been 
made". Assuming the technical difficulties involved in making this 
statement precise can be overcome, the Zurek-Thorne interpretation uses 
an ensemble of black holes and thus simply accepts the universality of 
$S_{BH}$ without offering any dynamical explanation for how it arises in 
a particular black hole. \par
No direct insight into the statistical origins of black hole entropy can come 
from thermodynamics. But if entropy really is a meaningful state function 
for black hole equilibrium states, then thermodynamics can tell us its value 
and provide an operational definition of it. To find the entropy of any 
thermodynamical state one invents a reversible process which arrives at 
that state from a state of known entropy and computes how the entropy 
changes in that process using the first law of thermodynamics. In this 
paper we examine the reversible contraction of a thin spherical shell down 
to its event horizon. To maintain reversibility, the shell must be in 
equilibrium with the acceleration radiation seen by observers on the shell. 
In addition to the classical stress-energy outside the shell there will be a 
Boulware stress-energy created by the quantum fields that live in the 
spacetime. To maintain thermal equilibrium we draw on a source of 
energy at infinity to ``top up" the Boulware stress-energy to an appropriate 
thermal environment. The pressure and surface density of the shell follow 
from the junction conditions at the shell. Using the first law we find that 
the entropy of the shell is ${A/4}$ in the black hole limit for a large class 
of shell equations of state. These considerations lead us to suggest the 
following operational definition for the entropy of a black hole: $S_{BH}$ 
is the equilibrium thermodynamic entropy that would be stored in the 
material which gathers to form the black hole, if we imagine all of this 
material compressed into a thin layer near its gravitational radius. 

\section{The Entropy of a Contracting Shell}

In this section we consider compressing a spherical shell reversibly from 
an infinite radius down to its event horizon. To maintain reversibility at 
each stage the shell must be in equilibrium with the acceleration radiation 
that would be measured by an observer on the shell. Thus the temperature 
of the shell is determined by the local acceleration of static observers at the 
shell. Our interest is in the end-states of a thermodynamic process: a state 
of infinite dispersion at infinity and the final black hole state. The shell 
serves merely as the working substance connecting these two states, and 
the nature of the material is irrelevant so long as it satisfies the first law of 
thermodynamics. Its equation of state involves two independent variables: 
the shell's locally measured mass $M$ and radius $R$. \par
For the static spherical geometries inside and outside the shell it will be 
general enough for our purposes to take the metric to be of the form

\begin{equation}\label{metric}
ds^2 = dr^2/f(r)+r^2d\Omega^2-f(r)dt^2.
\end{equation}

This covers as special cases Minkowski, Schwarzschild, Reissner-
Nordstr\"{o}m and de Sitter spacetimes. Of course, the classical stress-
energy associated (via the Einstein equations) with this metric is not the 
stress-energy of the ground state for the quantum fields which live in the 
spacetime. We know that this is the Boulware state \cite{fulling}, whose 
stress-energy $(T_{\mu\nu})_B$ depends on the types and number of 
fields and is unknown. For an ordinary star $(T_{\mu\nu})_B$ is 
completely negligible, but as the shell approaches its gravitational radius it 
generally grows without bound, and its back-reaction cannot be ignored. 
We cannot compute this back-reaction, but we can compensate for it. By 
drawing from an energy reservoir at infinity we fill up the shell's 
surroundings with material whose stress-energy $\Delta T_{\mu\nu}$ tops 
up $(T_{\mu\nu})_B$ to form a thermal bath which shares the shell's 
local acceleration temperature $T_{acc}(R)$ at the point of contact.
This ``topped-up Boulware 
state" (TUB) is constructed in thermal quantum field theory by 
periodically identifying the co-ordinate $t$ in the Euclidean sector with 
period equal to the reciprocal of the shell's redshifted acceleration 
temperature $T_{\infty}=T_{acc}(R)f(R)^{1/2}$. Then the TUB's local 
temperature varies in accordance with Tolman's law \cite {tolman}

\begin{equation}\label{tolman}
T(r)(-g_{tt})^{1/2}=T_{\infty}=constant.
\end{equation}

The TUB may be called a generalized Hartle-Hawking (HH) state. Indeed, 
it becomes the HH state in the limit when the shell approaches its 
gravitational radius. Its stress-energy

\begin{equation}
(T_{\mu\nu})_{TUB}=(T_{\mu\nu})_B + \Delta T_{\mu\nu}
\end{equation}

is, like the HH stress-energy, everywhere bounded and small for a large 
black hole, but non-vanishing at infinity. \par
To keep effects of back-reaction under control, we encase the TUB in a 
large spherical container of radius $R_{big}$. Back-reaction is negligible 
if the total energy of the TUB is small compared to the shell's mass $M$, 
i.e., (in Planck units)

\begin{equation}
T_\infty^4 R_{big}^3 \ll M,
\end{equation}

or in conventional units,

\begin{equation}
R_{big}/(2GM/c^2) \ll (M/m_{pl})^{2/3} \approx 10^{25} 
(M/M_{\odot})^{2/3}.
\end{equation}

We assume this condition satisfied, and we shall ignore back-reaction and 
also the entropic contribution of the TUB. \par
Phenomenology gives us the freedom of a dualistic approach. The thermal 
equilibrium condition $T_{shell}=T_{TUB}$ corresponds to the 
viewpoint of a local \emph{stationary} observer. On the other hand, for 
the stress-energy of the TUB we adopt the ``objective" (gravitating) value 
which appears on the right-hand side of the Einstein equations and 
corresponds to what is measured by a local \emph{free-falling} observer, 
and we take this (and the associated TUB energy) to be negligible for a 
large black hole. The TUB would look very different to a stationary 
observer, for whom the ground state is the Boulware state. In a statistical 
analysis of the problem, this observer's view is the one 
we would be forced to adopt, 
since no technique is currently available for analyzing the statistical 
thermodynamics of a system in anything other than its stationary rest-
frame. Such an analysis would lead to values for the TUB's apparent 
energy and entropy which are large, and divergent in the black hole limit. 
The extensive literature devoted to this problem resorts to various 
procedures (e.g. ``brick-wall" cutoffs \cite{thooft}, renormalization of the 
gravitational coupling constant \cite{susskind}) to tame these divergences. 
\par 
Now consider the surface stress-energy of the shell. The interior 
and exterior metric will be of the form (\ref{metric}) with
$f(r)=f_1(r)$ and $f=f_2(r)$ respectively. 
The surface stress-energy is 
related to the ``jump" in the extrinsic curvature via \cite{shell}

\begin{equation}
8\pi S_{ab}=[K_{ab}-g_{ab}K]
\end{equation} 

where $K_{ab}$ is the extrinsic curvature and $[\cdots]$ denotes the jump 
in the quantity in brackets (Latin indices, $a,b$, etc. run from 1 to 3). A 
simple calculation gives

\begin{equation}
4\pi \sigma = -\left[\sqrt{f(R)}/R\right]
\end{equation}

and

\begin{equation}\label{P}
16\pi P=\left[f'/\sqrt{f}+2\sqrt{f}/R\right]
\end{equation}

where $\sigma$ is the proper surface density of the shell and $P$ is the 
surface pressure. Since the mass of the shell as seen by local free-falling 
observers is $M=4\pi \sigma R^2$ we have 

\begin{equation}\label{M}
M=-\left[R\sqrt{f}\right].
\end{equation}

It is instructive to see the explicit form of these expressions for a shell of 
charge $e$ and gravitational mass $m$ with a flat interior. We therefore 
set $f_1(r)=1$ and $f_2(r)=1-2m/r+e^2/r^2$ in (\ref{P}) and (\ref{M}) 
and find that 

\begin{equation}\label{MP}
m=M-\frac{1}{2}\left( \frac{M^2-e^2}{R} \right) \ \ \mbox{and}\ \  
P=\frac{M^2-e^2}{16\pi R^2(R-M)}.
\end{equation}

These expressions have obvious Newtonian counterparts and simple 
intuitive meanings. \par
As discussed above, if the shell is to be contracted reversibly it must be in 
equilibrium with the acceleration radiation that would be seen by 
observers on the shell. Thus the shell's temperature must be given by 
\cite{unruh}

\begin{equation}
T=a/2\pi=f'(R)/4\pi\sqrt{f(R)}.
\end{equation} 

For a Reissner-Nordstr\"{o}m space-time $f(r)=1-2m/r+e^2/r^2$ and

\begin{equation}
T=\frac{2MR-M^2-e^2}{4\pi R^2(R-M)}.
\end{equation}

Since the local gravitational acceleration is discontinuous across the shell, 
the inner and outer TUBs in which the shell is immersed are at different 
temperatures. To maintain equilibrium an ``adiabatic" diaphragm 
(impermeable to heat) must be interposed between the faces. We can 
picture the shell as a pair of concentric spherical plates, with inner and 
outer masses $M_1$ and $M_2$, separated by a massless and thermally 
inert interstitial layer of negligible thickness. How we distribute the total 
shell mass $M=M_1+M_2$ between the plates is arbitrary. We choose 
$M_1$ so that the spacetime is flat between the plates. This generally 
makes $M_1$ negative. The two plates thus separate three concentric 
spherical zones: an inner zone where $f(r)=f_1(r)$, a very thin 
intermediate zone where $f(r)=1$ and an outer zone where $f(r)=f_2(r)$. 
Applying (\ref{P}) and (\ref{M}) to the inner and outer plates gives for the 
masses $M_i$ and surface pressures $P_i$ (i=1,2)

\begin{equation}\label{mipi}
M_i=R\xi_i (1-V_i(R))\ \ \mbox{and} \ \ 16\pi P_i=\left(\frac{\xi_i 
f'_i(R)}{V_i(R)}-\frac{2 M_i}{R^2}\right),
\end{equation}

where $f_i$ and $V_i=f_i^{1/2}$ are evaluated at $r=R$, the common 
radius of the two plates, and $\xi_i=(-1)^i$. The temperature $T_i$ of the 
plates is given by

\begin{equation}\label{ti}
T_i=\frac{f_i'(R)}{4\pi V_i(R)}.
\end{equation}

This gives $T_i$ as a function of $M_i$ and $R$.\par
Now consider the first law of thermodynamics, which would usually relate 
$dS$ to the quantity $(dM+PdA)/T$ in terms of the variables discussed so 
far. But since we are using an explicit function $T_i(M_i,R)$, the quantity 
$(dM+PdA)/T$ will not in general be an exact differential and hence it 
cannot be a complete representation of the differential $dS$. Thus we need 
to introduce another thermodynamic variable $N=N(M,R)$. Since the 
plates are merely abstract entropy-carrying devices, the physical 
significance of $N$ is irrelevant. For convenience we interpret $N$ as the 
number of particles in the shell. The first law now becomes (temporarily 
dropping the index $i$)

\begin{equation}
dS=\beta dM +\beta P dA - \alpha dN,
\end{equation}

where $\beta = 1/T$, $\alpha=\mu/T$, and $\mu$ is the chemical potential. 
Using the Gibbs-Duhem relation

\begin{equation}\label{gd}
S=\beta(M+PA)-\alpha N
\end{equation}

gives

\begin{equation}
n d\alpha=\beta dP + (\sigma+P)d\beta, 
\end{equation}

where $n=N/A$. Using the formulae (\ref{ti}) and (\ref{mipi}) for $T$,  
$P$ and $\sigma$ in terms of $M$ and $R$ gives

\begin{equation}\label{nda}
nd\alpha=\frac{\sigma}{\gamma}d\left(\frac{1}{2\xi\sigma\gamma}\right
),
\end{equation}

where

\begin{equation}\label{gamma}
\gamma^2=\frac{f'}{8\pi \xi\sigma V}.
\end{equation}

The functions $n$ and $\alpha$ can be chosen arbitrarily subject only to 
the restriction imposed by (\ref{nda}). The simplest option is to choose 
plate materials having the ``canonical" equation of state (denoted by an 
asterisk, and restoring the index $i$)

\begin{equation}
n_i^{\ast}=\frac{\sigma_i}{\gamma_i}\ \ \mbox{and} \ \ \alpha_i^{\ast} = 
(2\xi_i\sigma_i\gamma_i)^{-1}.
\end{equation}

Now, from (\ref{ti}) and (\ref{gamma}) we have

\begin{equation}\label{ti2}
T_i=2\xi_i\sigma_i\gamma_i^2.
\end{equation}

Thus the canonical chemical potential $\mu_i^\ast = T_i \alpha_i^\ast$ 
obeys the simple relation

\begin{equation}
\mu_i^\ast n_i^\ast = \sigma_i.
\end{equation}

Substituting the above into (\ref{gd}) and noting from (\ref{mipi}) and 
(\ref{gamma}) that the surface pressures can be expressed as

\begin{equation}
P_i=\frac{1}{2}\sigma_i (\gamma_i^2 -1),
\end{equation}

we obtain the entropy density $s_i^\ast=S_i^\ast/A$ of the plates as

\begin{equation}\label{s}
s_i^\ast=\beta_i P_i=\frac{1}{4}\xi_i(1-\gamma_i^{-2}).
\end{equation}

When the slow contraction of the shell terminates it is hovering just 
outside the horizon ($r=r_0$, defined by $f(r_0)=0$) of the exterior 
geometry. Now consider the non-extremal case where the surface gravity 
$\kappa=f_2'(r_0)/2 \neq 0$ (the extremal case will be examined in 
section 5). Then (\ref{gamma}) shows that $\gamma_2^2$ diverges 
according to 

\begin{equation}\label{div}
\gamma_2^2 \approx \frac{\kappa_2}{M_2/R^2}V_2^{-1}\ \mbox{with} 
\ V_2^2=2\kappa(R-r_0)  
\end{equation}

as $R\to r_0$. Thus, from (\ref{s})

\begin{equation}\label{slim}
\lim_{R \to r_0} s_2^\ast=\frac{1}{4}.
\end{equation}

That is, \emph{the entropy of the outer plate is one-quarter of its area in 
Planck units in the black hole limit}. In the simplest situation the spherical 
cavity inside the shell is flat and empty. In this case $f_1(r)=1, 
M_1=P_1=s_1=T_1=0$ and the outer plate contributes all of the mass and 
entropy of the shell. 

\section {Black Hole Entropy}
In the previous section we found that the entropy of a shell with a flat 
interior is one quarter of its area in the black hole limit. From an 
observer's perspective at infinity there is nothing to distinguish the shell in 
its final stages of compression from a black hole. We could even arrange 
for the shell to leak out energy and entropy in a simulated Hawking 
evaporation. This suggests an operational definition for the entropy of a 
black hole, namely the limiting entropy of the associated shell. \par
But is this definition of entropy additive? Suppose that, in the field of a 
pre-existing black hole with Bekenstein-Hawking entropy $S_{old}$ (or 
of any object, e.g., a star, having this entropy), we lower a shell of entropy 
$S_{shell}=S_2+S_1$ to the point where an outer black hole, of area 
$A_{new}$, is about to form, so that $S_2=A_{new}/4$ for the outer 
plate according to (\ref{slim}). Is the new Bekenstein-Hawking entropy 
obtained by simple addition as $S_{old}+S_{shell}$? At this point 
certainly not. The upper plate by itself already accounts for the full 
Bekenstein-Hawking entropy of the new configuration, so it would be 
necessary for the negative entropy of the inner plate to cancel exactly the 
entropy of whatever was inside the cavity initially, i.e. $S_1+S_{old}$ 
would need to be $0$. This is generally not true. However, we are still free 
to carry out a further reversible procedure: we can sweep all the material 
inside the cavity onto the outer shell. (If this material includes an inner 
black hole, this involves inflating the shell representing it until it merges 
with the lower plate of the new shell - in effect, a (reversible) 
``evaporation" of the inner black hole). This ``flattens" the cavity inside 
the new shell and dematerializes the lower plate. ( In a more general 
(rotating) context, evacuation will not flatten the cavity, but could still 
reduce the acceleration temperature at the lower plate to zero.) With the 
shell's entropy thus modified, the total entropy at the end is 
$S_{shell}'=A_{new}/4$, which is $S_{BH}$ for the final black hole. In 
this specific sense, $S_{BH}$ may be called ``additive", but it is perhaps 
more correct to say it is ``forgetful": $S_{BH}$ for the final configuration 
betrays no clue about the entropy originally contained in 
the space now occupied by the hole. \par 
These considerations suggest the following operational definition: \emph{ 
$S_{BH}$ is the equilibrium thermodynamic entropy that would be stored 
in the material which gathers to form the black hole, if all of this material 
were compressed into a thin layer near its gravitational radius}. Since the 
entropy for a given mass and area is maximized for thermal equilibrium 
we expect that this is the maximum entropy that could be stored in the 
material before it crosses the horizon. \par
Of course, this imagined process bears no resemblance to any real scenario 
of black hole formation. But as mentioned it can give a fair schematic 
description of the evaporation process since Hawking's mechanism of 
virtual pair creation is a skin effect confined to a thin layer near the 
horizon. In the real process, the horizon is a port where gravity temporarily 
detains the evaporating particles on their way out of the hole; the shell 
model assembles all of them there at one time. Kundt's description of 
$S_{BH}$ as ``evaporation entropy" sums up the situation rather well, 
with the proviso that the evaporation process itself (being virtually 
reversible) cannot be the \emph{source} of $S_{BH}$; it only acts as its 
conduit. \par

\section{Alternative Plate Material}

The key result (\ref{slim}) was established for a special ``canonical" form 
of the plate material. How sensitive are the results to the properties of the 
material? The most general functions $n$ and $\alpha$ satisfying 
(\ref{nda}) are obtained by replacing $\alpha_i^\ast$ by an arbitrary 
function of itself $g_i(\alpha_i^\ast)$, and replacing $n_i^\ast$ with 
$n_i^\ast / g'_i(\alpha_i^\ast)$. This yields the general formulas 

\begin{equation}\label{an}
\alpha_i=g_i(\alpha_i^\ast), \ \ n_i=n_i^\ast/g'_i(\alpha_i^\ast)
\end{equation} 

and

\begin{equation}
\frac{\mu_i n_i}{\sigma_i} = \frac{g_i}{\alpha_i^\ast g'_i}.
\end{equation}

The most general expression for the entropy density of the plates is 

\begin{equation}\label{sgen}
s_i=\frac{1}{4}\xi_i[1+\gamma^{-2}(1-2\mu_i n_i/\sigma_i)].
\end{equation}

Thus, (\ref{slim}) is invariant under arbitrary transformations of the form 
(\ref{an}) which leave $\mu_i n_i$ bounded in the high temperature limit. 
Indeed, we can allow transformations which are singular in this limit, 
provided that

\begin{equation}\label{const}
\lim_{T_i \to \infty } \frac{\mu_i n_i}{T_i}=0,
\end{equation}

recalling (\ref{ti2}). Expression (\ref{slim}) is not invariant 
under arbitrary singular transformations. At the root of this problem is the 
fact that the black hole end-state is a singular state of the plate material 
($P$ and $T$ become infinite). In these circumstances there is no a priori 
justification for excluding or constraining asymptotically singular behavior 
of thermodynamic quantities. However, it is reassuring to note that the 
loose constraint (\ref{const}) guarantees that our conclusions are 
independent of the plate material for a very broad class of equations of 
state. \par
The freedom contained in the transformations (\ref{an}) can be used to 
``improve" the behavior of the plate material at low temperatures. For 
canonical material the total entropy (not the entropy density) goes as 
$S^\ast_{i(tot)} \approx R$, ($R \to \infty$), since $\gamma^2 \approx 1-
M/R$ for $R \to \infty$. By a suitable change of $g_i$ in (\ref{an}) we can 
arrange that $S^\ast_{i(tot)}$ is finite in the limit $R \to \infty$ (and $T_i 
\to 0$).

\section{Extremal Black Holes and the Third Law}

There are essentially two distinct versions of the third law of 
thermodynamics. The first version, proposed by Nernst in 1906, states that 
isothermal processes become isentropic in the zero temperature limit.  An 
essentially equivalent form states that the temperature of a system cannot 
be reduced to zero in a finite number of operations. The second version, 
proposed by Planck in 1911, states that the entropy of any system tends, as 
$T \to 0$, to an absolute constant, which may be taken as zero. \par
In their 1973 paper on ``the four laws of black hole dynamics", Bardeen, 
Carter and Hawking \cite{bardeen} proposed a form of the third law 
patterned after Nernst's unattainability principle: ``It is impossible by any 
process, no matter how idealized, to reduce the surface gravity to zero in a 
finite sequence of operations." A more specific form \cite{israel}, which 
makes precise the meaning of ``a finite sequence of operations" states: ``a 
non-extremal black hole cannot become extremal at finite advanced time 
in any continuous process in which the stress-energy of accreted matter 
stays bounded and satisfies the weak energy condition." From this 
formulation it is clear that quantum processes like evaporation, which 
typically involve the absorption of negative energy, can violate Nernst's 
form of the third law. \par
For a long time it was believed that there is no black hole analogue to 
Planck's version of the third law. Recently, however, this has become a 
matter of controversy. Arguments based on black hole instanton topology 
and pair creation \cite{hhr} suggest that the entropy of extremal black 
holes is zero. On the other hand, the remarkable indirect calculations of 
black hole entropy by counting states of strings on D-branes gives the 
value $S_{BH}=A/4$ for extremal black holes \cite{10}. \par 

We can examine this question by considering the quasistatic contraction of 
an extremally charged spherical shell with a flat interior. Setting $|e|=m$ 
in (\ref{M}) and (\ref{MP}) we find that $M=|e|$ and $P=0$. If the shell 
is made of canonical material (\ref{s}) gives $s^\ast=0$ at all stages of the 
contraction, leading to 

\begin{equation}
S_{extBH}=0
\end{equation}

This result is, however, quite sensitive to the equation of state of the shell 
material. For arbitrary material, (\ref{sgen}) gives the shell's entropy 
density as

\begin{equation}
s_{extBH}=\frac{1}{2}\left(1-\frac{\mu n}{\sigma}\right)
\end{equation}

whose value can be made arbitrary by choice of the function 
$g(\alpha^\ast)$ in (\ref{an}). Thus no universal quantity can be assigned 
to the entropy of an extremal shell in any stage of its compression. This 
suggests that the entropy of extremal black holes may depend on their 
prior history.\par

\section{Conclusion}
In this paper we calculated the entropy of a quasistatically contracting 
spherical shell and discussed its relationship to the entropy of the black 
hole that it forms. \par
Outside the shell the classical stress-energy tensor will be modified by the 
quantum fields that live in the spacetime. The Boulware stress-energy 
produced by these fields was ``topped up" to provide a thermal 
environment and reduce the back-reaction to negligible levels for large 
black holes. Since the shell contracts reversibly its temperature 
must be equal to the acceleration temperature seen by observers on 
the shell. The surface pressure and density follow from the 
junction conditions. We reformulated the Gibbs relation in the form 
(\ref{nda}), involving $n=N/A$ and $\alpha=\mu/T$ where $N$ is the 
number of particles, $A$ the area, $\mu$ the chemical potential and $T$ 
the temperature of the shell. The entropy of the shell can easily be 
calculated once $n$ and $\alpha$ are found. The solution to the equation 
for $n$ and $\alpha$ is not unique, but we found a simple solution (the 
``canonical" solution). In the non-extremal case the entropy of a shell 
made of canonical material approaches $A/4$ as the shell approaches its 
event horizon. This result does not hold for all solutions for $\alpha$ and 
$n$. However, it does hold for all equations of state which satisfy $\mu 
n/T \to 0$ as the shell approaches its event horizon (note that $T \to \infty$ 
in this limit).\par
The considerations in this paper led us to suggest the following operational 
definition for the entropy of a black hole: $S_{BH}$ is the equilibrium 
thermodynamic entropy that would be stored in the material which gathers 
to form the black hole, if all of this material were compressed into a thin 
layer near its gravitational radius. Since the entropy for a given mass and 
area is maximized for thermal equilibrium we expect that this is the 
maximum entropy that could be stored in the material before it crosses the 
horizon. \par
For the special case of an extremal shell (charge equals mass in relativistic 
units) our approach gives ambiguous results; the limiting entropy of the 
shell depends on the equation of state of the material. \par
It should be noted that these conclusions go significantly beyond the 
verifications of the generalized second law 
(\cite{generalized},\cite{zurek}) which show that one-quarter of the 
\emph{change} in area (when a black hole slowly ingests material) is 
equal to the entropy absorbed. Here, we have derived the entropy-area 
relation in integral form, eliminating the possibility of an additive 
constant. \par
\vfill
\noindent
{\bf Acknowledgements} We would like to thank Warren Anderson for 
many stimulating discussions.\par
This work was supported by NSERC of Canada and by the Canadian 
Institute for Advanced Research.

\end{document}